\documentclass[journal]{IEEEtran}
\usepackage{cite}
\usepackage{amsmath,amssymb,amsfonts}
\usepackage{commath}
\usepackage{algorithmic}
\usepackage{graphicx}
\usepackage{textcomp}
\usepackage{xcolor}
\usepackage{float}
\usepackage{hyperref}
\usepackage{soul}
\usepackage{siunitx}

\usepackage[linesnumbered,ruled,vlined]{algorithm2e}

\usepackage{amsmath}
\usepackage{setspace}
\usepackage{color}

\begin{document}

\pagenumbering{arabic}

\title{Analysis and Design of Analog Fountain Codes for Short Packet Communications}

\author{\IEEEauthorblockN{Wen Jun Lim,~\IEEEmembership{Graduate Student Member,~IEEE}, Rana Abbas,~\IEEEmembership{Member,~IEEE}, Yonghui Li,~\IEEEmembership{Fellow,~IEEE}, 
Branka Vucetic,~\IEEEmembership{Life Fellow,~IEEE}, Mahyar Shirvanimoghaddam,~\IEEEmembership{Senior Member,~IEEE}}

\thanks{Manuscript received May 26, 2021; revised August 24, 2021, and accepted September 30, 2021. Date of publication October ..., 2021; date of current version October ..., 2021. This work was supported by the Australian Research Council through the Discovery Projects under Grants DP180100606 and DP190101988. 

This paper was presented in part at the IEEE 90th Vehicular Technology Conference, Honolulu, HI, USA, Sep. 2019. \textit{(Corresponding author: Wen Jun Lim.)}

The authors are with the Centre for IoT and Telecommunications, School of Electrical and Information Engineering, The University of Sydney, NSW 2006, Australia (e-mail: \{wenjun.lim; rana.abbas; yonghui.li; branka.vucetic; mahyar.shirvanimoghaddam\}@sydney.edu.au).

This paper has been accepted for publication by IEEE. Copyright (c) 2017 IEEE. Personal use of this material is permitted.  However, permission to use this material for any other purposes must be obtained from the IEEE by sending a request to pubs-permissions@ieee.org.}}

\maketitle

\begin{abstract}
In this paper, we focus on the design and analysis of the Analog Fountain Code (AFC) for short packet communications. We first propose a density evolution (DE) based framework, which tracks the evolution of the probability density function of the messages exchanged between variable and check nodes of AFC in the belief propagation decoder. Using the proposed DE framework, we formulate an optimisation problem to find the optimal AFC code parameters, including the weight-set, which minimises the bit error rate at a given signal-to-noise ratio (SNR). Our results show the superiority of our AFC code design compared to existing designs of AFC in the literature and thus the validity of the proposed DE framework in the asymptotically long block length regime. We then focus on selecting the precoder to improve the performance of AFC at short block lengths. Simulation results show that lower precode rates obtain better realised rates over a wide SNR range for short information block lengths. We also discuss the complexity of the AFC decoder and propose a threshold-based decoder to reduce the complexity.
\end{abstract}


\section{Introduction}
\IEEEPARstart{T}{he} third generation partnership project (3GPP) has defined three primary service categories for the fifth generation (5G) of mobile communications \cite{ghosh20195g}.  These include enhanced mobile broadband (eMBB), ultra-reliable and low-latency communications (URLLC), and massive machine type communications (mMTC). 5G aims at providing a higher data rate (e.g., up to 10 Gbps for eMBB), shorter end-to-end latency (e.g., $\le$ 1 ms for URLLC), higher reliability (e.g., packet error rate (PER) of less than $10^{-5}$ for URLLC) and higher energy efficiency (e.g., up to 10 years operation with a single battery for mMTC) \cite{arora2019survey}. Many new techniques have been proposed to meet these requirements, e.g., massive multiple-input-multiple-output (mMIMO), millimetre wave (mmW) communications, non-orthogonal multiple access (NOMA), short-packet communications, etc. \cite{li20185g,mei2019survey}. For example, the authors in \cite{sun2018short}  demonstrated that NOMA has superior performance in terms of fairness and spectral efficiency compared to orthogonal multiple access (OMA) for low latency downlink in the short packet scenario. In \cite{li2018channel}, the authors derived the expression for the minimum channel reciprocity that enables the uplink channel training to achieve a higher data rate than the downlink channel training. Results showed that the advantage of uplink channel training in short packet communication with multiple transmit antennas as the channel reciprocity coefficient decreases with the decrease of the block length or the increase of the number of transmit antennas. The authors in \cite{li2019optimal} proposed  the optimal resource allocation to maximize the average data rate in the MIMO systems, which adopts short packet communications. With the emergence of new applications and services, such as extreme URLLC, and entering the era of Beyond 5G (B5G) and eventually moving towards the 6th generation (6G) of wireless communications standard, fundamental breakthroughs in the way we design communications systems are vital in order to meet the requirements of emerging applications. 

The channel code design has been always at the heart of any communications standards and still remains an essential part to unleash the full potential of 5G New Radio (NR) services, namely, to guarantee the required latency and reliability requirements at maximum efficiency \cite{arora2019survey}. It has been shown that by using shorter transmission-time interval (TTI), lower latency can be achieved, which implies using short packets \cite{shirvanimoghaddam2018short}. This is of particular importance for resource-constraint settings, where the available bandwidth, power, and hardware resources, such as antennas, are limited. The design of efficient and robust short packet communications is vital for 5G mMTC and URLLC applications, where the data transferred might be significantly smaller than traditional human-based communications, ranging from a few hundred bits down to a few tens \cite{wang2019secure,schulz2017latency}. Modern channel coding techniques have diverged away from the traditional theories of Shannon due to this new shift alongside other challenges imposed by the 5G services, e.g., higher reliability requirements (lower than $10^{-5}$) and higher energy efficiency. We refer the readers to the survey paper \cite{shirvanimoghaddam2018short} for a comprehensive review of modern channel coding techniques. In particular, there is a large consensus that traditional channel codes, adopted in current cellular networks, are strictly sub-optimal for short packet communications \cite{shirvanimoghaddam2018short}.

The conventional channel coding methods cannot fulfill the requirements of B5G and future 6G, such as large coding gain, low-complexity, low-latency, high-throughput, bit-level granularity, and flexible code parameters \cite{zhao2019adaptive}. Flexibility is required to provide different levels of reliability and latency at various block lengths. On the other hand, bit-level granularity implies ease of generating various rates and lengths with negligible performance degradation. The conventional adaptive modulation and coding which relies on the channel state information (CSI) feedback from the receiver is not suitable for short packet communications due to the significant overhead. Moreover, the signalling is estimated to incur 5–8 ms latency which violates the low latency requirement of URLLC \cite{abbas2019novel}. self-adaptive channel codes can be effectively used to reduce the end-to-end latency without any CSI at the transmitter side. Existing self-adaptive codes, also known as rateless codes, are either binary in nature with no straight-forward extension to adaptive modulation, or they exhibit significant performance degradation in the finite block length regime. In this paper, we focus on the design of efficient self-adaptive codes for short packet communications, i.e., for message lengths from only a few ($\le$100) bits up to a few hundred ($\le$1000) bits.

\subsection{Related Works}
The original research on rateless codes, such as LT codes \cite{luby2002lt} and Raptor codes \cite{shokrollahi2006raptor}, have shown that rateless codes can achieve the channel capacity over binary erasure channels. However, all design extensions of rateless codes to noisy channels, e.g., additive white Gaussian noise (AWGN) channel, have shown to be channel dependent. To the best of our knowledge, only a few rateless codes to date are near capacity-achieving over a wide range of signal to noise ratios (SNRs) for an asymptotically long block length. This includes Analog Fountain Codes (AFC) \cite{shirvanimoghaddam2013near,shirvanimoghaddam2013adaptive} and Spinal codes \cite{perry2011rateless}. These codes, however, show a significant gap to the the capacity at short block lengths \cite{wang2019secure,8336882}. 

In rateless codes, the rate is determined on the fly and without the need for CSI at the transmitter side. This is particularly favourable in fast varying channels as well as non-reciprocal channels. Recent studies on short AFC \cite{abbas2019novel} have shown impressive results for 5G URLLC in terms of comparable latency to the Polyanskiy-Poor and Verdu (PPV) normal approximation \cite{polyanskiy2010channel}, as well as reliability down to $10^{-7}$. However, the proposed encoding scheme therein, as well as the weight-set design, are based on heuristics rather than a solid analytical framework. In this paper, we aim to bridge this gap in the literature by providing a solid analytical framework and an optimisation framework for AFC. To the best of our knowledge, the only work that has attempted to do so can be found in \cite{zhang2020finite}. Authors in \cite{zhang2020finite} proposed a modified weight selection scheme for short block length AFC. In particular, the extrinsic information transfer (EXIT) chart analysis \cite{ashikhmin2004extrinsic} was modified in \cite{zhang2020finite} to address the issue of performance degradation of EXIT analysis in the finite block length regime. However, the design in \cite{zhang2020finite} relies on approximate densities and the assumption of symmetric Gaussian distribution for the messages exchanged in each iteration of the belief propagation (BP) decoder. These assumptions are not accurate \cite{jayasooriya2017analysis}, particularly at low rates. This negatively impacts the optimal code ensembles' search.

The authors in \cite{rao2018efficient,lu2017parallel} proposed a weight selection strategy according to channel condition, for reduced complexity, without loss of performance. Several other works mainly focused on the decoder design, where different approaches were proposed to reduce the complexity of the message passing decoder of AFC  \cite{lu2018low,rao2013log}. The design of AFC weight set in all these works were mostly heuristic and did not provide meaningful  insights into the code design for short block lengths. Furthermore, the performance of AFC at short block lengths depends on the precoder. However, the choice of the precoder for AFC has not been studied previously. 

\subsection{Main Contributions}
Our main contributions in this work are summarized next.

\subsubsection{A Density Evolution-based Framework to Analyse AFC}
A density evolution (DE) framework is proposed to analyse the performance of the BP decoder of AFC in the infinite block length regime. We track the evolution of $\mathrm{pdf}$ of log-likelihood ratios (LLR), which are exchanged between variable and check nodes during the BP decoding.  Due to the complexity of the variable and check node updating rules, we propose a Monte-Carlo based analysis to derive the $\mathrm{pdf}$s in each iteration of DE. Our results show that the proposed DE framework can precisely approximate the $\mathrm{pdf}$ of LLRs in each iteration of the BP decoding.

\subsubsection{A Differential Evolution Algorithm to Optimise AFC parameters} 
We use differential evolution to find the optimal weight set of AFC with different degrees by using the proposed DE framework. Simulation results show that by applying the optimised weight set, the performance of AFC in terms of bit error rate is improved compared with the existing designs of AFC in the literature, thus proving the validity of the proposed DE framework. 

\subsubsection{Precode selection for AFC in the short block length regime}
The DE framework proposed in this work is valid for the asymptotically long block length. To guarantee the performance at short block lengths, one needs to choose a proper precoder. We will show that the optimised weight set obtained from the DE framework in the asymptotic regime can be directly used in the short block length regime with the help of a precode with a powerful error-correcting capability to achieve the desired level of reliability. 
In particular, we consider Bose, Chaudhuri, and Hocquenghem (BCH) codes with an ordered statistics decoder (OSD) \cite{Fossorier1995} and LDPC codes with the BP decoder. We show that low rate precoders offer better reliability at a wide range of SNRs compared to high-rate precoders. We also show that the precoded AFC performs close to the normal approximation benchmark \cite{polyanskiy2010channel} in the finite block length regime over a wide range of SNRs. 

\subsubsection{Threshold-based Decoder for reducing the decoding complexity} We propose a Threshold-based decoder for BCH precoded AFC,  where only the soft information from AFC decoder, which deemed to be decoded with high confidence, will be passed to the BCH decoder. This prevents to execute the OSD algorithm in each and every iteration, which lead to the reduced complexity. Our results shows that this approach significantly reduces the decoding complexity with a negligible performance degradation.

\subsection{Paper Organisation}
The rest of the paper is organised as follows. In Section II, we explain the encoder and decoder of the precoded AFC code. We then propose the density evolution framework for the analysis and design of AFC in Section III. Section IV presents the differential evolution optimisation framework for the AFC weight-set optimisation. We then study the precode selection for AFC codes in the short block length regime in Section V. The decoding complexity of AFC and the proposed threshold-based decoder are discussed in Section VI. Finally, conclusions are drawn in Section VII. 
\begin{table}[t]
\centering
\caption{Summary of notations.}
\label{table:3}
\scriptsize
  \begin{tabular}{|p{1.2cm}|p{6.7cm}|} 
 \hline
  \textbf{Notation} & \textbf{Description}\\ 
 \hline
 $\mathbf{b}$& The information block\\
 \hline
 $k$& The length of the information block, i.e., $|\mathbf{b}|=n$\\
 \hline
 $R_{\mathrm{pre}}$& The rate of the precoder\\
 \hline
  $G_{\text{pre}}$ & Generator matrix of precode\\
\hline
 $\mathbf{u}$& The vector of intermediate symbols, i.e., $\mathbf{u}=G_{\mathrm{pre}}\mathbf{b}$\\
 \hline
 $n$& The number of intermediate symbols, i.e., $n=k/R_{\mathrm{pre}}$\\
 \hline
$\Omega(x)$ & Degree distribution function of AFC, i.e., $\Omega(x)=\sum_i \Omega_ix^i$\\
\hline
$\mathcal{W}$ & Weight set of AFC, i.e., $\mathcal{W}=\{w_1,\cdots,w_d\}$\\
\hline
$\mathcal{W}^{\pm}$&Set of all positive and negative weight coefficients, i.e., $\mathcal{W}^{\pm}=\left\{w,-w:~w\in\mathcal{W}\right\}$\\
\hline
$c_i$& The $i$th AFC coded symbol\\
\hline
$d_v$ & Degree of variable nodes in the bipartite graph of AFC\\
\hline
$d_c$ & Degree of check nodes in the bipartite graph of AFC\\
\hline
$R_{\mathrm{AFC}}$& Rate of AFC code truncated at length $m$, i.e., $R_{\mathrm{AFC}}={n}/{m}=d_c/d_v$.\\
\hline
$\mathcal{M}_c\backslash v$ & Set of variable nodes other than $v$ connected to check node $c$\\
\hline
$\mathcal{N}_v\backslash c$ & Set of check nodes other than $c$ connected to variable node $v$\\
\hline
$m_s$& Number of AFC coded symbols to perform a successful decoding\\
\hline
$\mathbb{E}[.]$ & Expectation operand\\
\hline
$\mathcal{R}$ & Realised rate of AFC, i.e., $\mathcal{R}=k/\mathbb{E}[m_s]$\\
\hline
$\delta$ & Number of AFC coded symbols sent between decoding attempts\\
\hline
$\sigma^2$ & AWGN variance\\
\hline
$\gamma$ & Signal-to-noise-ratio (SNR), i.e., $\gamma=1/\sigma^2$\\
\hline
$m_{c\rightarrow v}^{(\ell)}(w)$ & Message sent from check node $c$ to variable node $v$ along the edge with weight $w$ at the $\ell$th iteration of BP\\
\hline
$f^{(\ell)}_{cv}(w,m)$ & The $\mathrm{pdf}$ of $m^{(\ell)}_{c\rightarrow v}(w)$\\
\hline
$m_{v\rightarrow c}^{(\ell)}$ & Message sent from variable node $v$ to check node node $c$ at the $\ell$th iteration of BP\\
\hline
$f^{(\ell)}_{vc}(m)$ & The $\mathrm{pdf}$ of $m^{(\ell)}_{v\rightarrow c}$\\
\hline
 \end{tabular}
\end{table}

\section{Analog Fountain Codes}
AFC was originally proposed in \cite{shirvanimoghaddam2013near}, which is mainly characterised by a weight set, degree distribution function, and the message length. The code is rateless in nature and can generate a potentially limitless number of coded symbols, thus, achieving any desired rate on the fly. In what follows, we explain the encoding and decoding processes of AFC. Table \ref{table:3} summarizes notations commonly used in this paper. 

\subsection{The Encoder}
The precoded AFC is a concatenation of a fixed-rate precode and the AFC code. An information block of length $k$ bits, denoted by $\mathbf{b}$, is first encoded by using a fixed-rate ($n,k$) code (the precode) of rate $R_{\text{pre}}=k/n$, to generate a codeword of length $n$ bits, denoted by $\mathbf{u}$, referred to as \emph{intermediate symbols}. The generator matrix of the fixed-rate code is denoted by $G_{\text{pre}}$; therefore, we have:
\begin{align}
\mathbf{u}=G_{\text{pre}}\mathbf{b}.
\end{align}
The precode serves as the outer code for the precoded AFC code. Intermediate symbols are modulated by using a BPSK modulation to generate $n$ modulated symbols, $v_i$, that is 
\begin{align}
    v_i=(-1)^{u_i},~~ \text{for}~i=1,\cdots,n.
\end{align}

\begin{figure}[t]
    \centering
    \includegraphics[width=0.6\columnwidth]{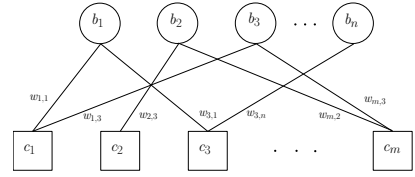}
    \caption{Bipartite graph representation of an AFC code.}
    \label{fig:fig10}
\end{figure}

Next, by using an AFC code a potentially limitless number of AFC coded symbols, also referred to as \emph{output symbols}, are generated. AFC is mainly characterised by a weight set $\mathcal{W}$ and a degree distribution function $\Omega(x)=\sum_{i=1}^{n}\Omega_ix^i$. In order to generate an AFC coded symbol, a degree ${d}$ is drawn based on the degree distribution function $\Omega(x)$.  Then ${d}$ modulated intermediate symbols are randomly selected and linearly combined in the real domain with a set of ${d}$ real weight coefficients selected from the weight set $\mathcal{W}$. For simplicity, we consider that the degree $d$ is fixed, i.e., $\Omega(x)=x^d$, and the weight set is predefined and given by $\mathcal{W}=\{w_1,w_2, \cdots, w_d\}$. The $i$th AFC coded symbol, denoted by $c_i$, is then given by: 

\begin{equation}
c_i= \sum_{j=1}^{d} w_{j} v_{i,j},
\label{eq:afcenc}
\end{equation}
where $v_{i,j}\in \mathcal{V}_i$ and $\mathcal{V}_i$ is the set of modulated intermediate symbols that have been chosen to generate the $i$th output symbol. We further assume that $\sum_{i=1}^{d}w_i^2=1$, therefore we have $\mathbb{E}[|c|^2]=1$, where $\mathbb{E}[.]$ is the expectation operand. Fig. \ref{fig:fig10} shows the bipartite graph representation of the AFC code truncated at length $m$. We refer to AFC coded symbols and intermediate symbols in the bipartite graph of AFC by check and variable nodes, respectively. The degree of a check (variable) node is defined as the number of variable (check) nodes connected to it in the bipartite graph. A regular AFC has constant variable and check node degrees, $d_v$ and $d_c$, respectively. As can be seen in Fig. \ref{fig:fig10} each edge of the graph is assigned a real weight from $\mathcal{W}$ and each check node is the real sum of the variable nodes multiplied by their weights assigned to their edges connecting them to that check node. 

AFC coded symbols are sent over the channel. Once the receiver received $m_0$ symbols, it performs the decoding. If the decoding failed, it collects $\delta$ additional AFC coded symbols and reruns the decoder. This process will continue until the decoding succeeds. The realised rate of the precoded AFC is given by 
\begin{align}
    \mathcal{R}=\frac{k}{\mathbb{E}[m_s]}=R_{\text{pre}}\times\frac{n}{\mathbb{E}[m_s]},
\end{align}
where $m_s$ is the number of AFC coded symbols collected to perform a successful decoding. 

\subsection{The Decoder}
We consider the additive Gaussian noise (AWGN) channel, where the channel output $y_i$ is given by
\begin{align}
    y_i=c_i+n_i,~~\text{for}~~i=1,2,\cdots,
\end{align}
where $n_i$ is AWGN with zero mean and variance $\sigma^2$. The signal-to-noise ratio (SNR), denoted by $\gamma$, is then given by $\gamma=1/\sigma^2$. 

The decoding is performed in two stages. First, the BP decoding algorithm is applied to AFC to find the LLRs of intermediate symbols. We use the BP algorithm initially proposed in \cite{baron2010bayesian} and further modified in \cite{cui2011seamless} to decode AFC. Second, the LLRs are passed to the decoder of the precode to find the original $k$ information symbols. If the decoding failed, the BP decoding is repeated with a longer block of AFC symbols which includes newly arrived symbols. The LLRs are passed again to the decoder of the precode. This continues until the decoding succeeded or the maximum number of AFC symbols are sent. 

In this paper, we consider both BCH codes and LDPC codes as the precoder. BCH code is a powerful cyclic error-correcting code with a variety of block lengths and corresponding code rates \cite{shirvanimoghaddam2018short}. BCH codes have strong error-correcting capability, which can correct all random patterns of $t$ errors, where $t$ is the design parameter. BCH codes are effective in preventing error floor due to the large minimum Hamming distance \cite{shirvanimoghaddam2018short}. However, there is a shortcoming of BCH codes, which is not flexible enough since block length and information length cannot be selected arbitrarily. For BCH precoded AFC, we use the ordered statistics decoder (OSD), which is computationally complex. LDPC, on the other hand,  offers a lower complexity decoder, as BP can be effectively used to decode them. However, LDPC cannot offer the same level of error correction capability as BCH codes, particularly at short block lengths.

\subsection{The Belief Propagation Decoding of AFC}
We consider a regular AFC with a constant variable and check node degree, $d_v$ and $d_c$, respectively. In each iteration of  the BP decoding, messages are exchanged between the check and variable nodes and vice versa. We use the LLR as the message which is exchanged between nodes in each iteration of the BP algorithm.

Let $m_{c\rightarrow v}^{(\ell)}(w_v)$ denote the message sent from check node $c$ to variable node $v$ along the edge with weight $w_v$ in the $\ell$th iteration of the BP decoding. It can be calculated as follows:
\begin{align}
\nonumber &m^{(\ell)}_{c\rightarrow v}(w_v)=\\
    &\ln \frac{\sum\limits_{\substack{b_{v'}\in \{-1,1\}\\v'\in\mathcal{M}_c\backslash v}}e^{-\frac{\left(y-w_v-\sum\limits_{v'\in\mathcal{M}_c\backslash v}w_{v'}b_{v'}\right)^2}{2\sigma^2}} \prod\limits_{v'\in\mathcal{M}_c\backslash v}p^{(\ell-1)}_{v'\rightarrow c}(b_{v'})}{\sum\limits_{\substack{b_{v'}\in \{-1,1\}\\v'\in\mathcal{M}_c\backslash v}}e^{-\frac{\left(y+w_v-\sum\limits_{v'\in\mathcal{M}_c\backslash v}w_{v'}b_{v'}\right)^2}{2\sigma^2}} \prod\limits_{v'\in\mathcal{M}_c\backslash v}p^{(\ell-1)}_{v'\rightarrow c}(b_{v'})},
    \label{eq:ctov}
\end{align}
where $y=c+n$ is the received signal corresponds to check node $c$, $\mathcal{M}_c\backslash v$ denote the set of variable node connected to check node $c$ except variable node $v$ and
\begin{align}p^{(\ell)}_{v'\rightarrow c}(b_{v'})=\left\{\begin{array}{ll}
\left(1+e^{-m^{(\ell)}_{v'\rightarrow c}}\right)^{-1},&~\text{if}~b_{v'}=1,\\
\left(1+e^{m^{(\ell)}_{v'\rightarrow c}}\right)^{-1},&~\text{if}~b_{v'}=-1,
\end{array}\right.
\label{eq:mtop}
\end{align}
and $m^{(\ell)}_{v\rightarrow c}$ denote be the message sent from a variable node $v$ to check node $c$ in the $\ell$th iteration of the BP decoding, which is given by
\begin{equation}
    m^{(\ell)}_{v\rightarrow c}=\sum_{c'\in \mathcal{N}_v\backslash c} m_{c'\rightarrow v}^{(\ell-1)}(w_{v'})
    \label{eq:vtoc}
\end{equation}
where $\mathcal{N}_v\backslash c$ represents the set of check nodes that are connected to variable node $v$ except check node $c$. It is important to note that as can be seen in \eqref{eq:ctov}, the message passed from a check node to a variable node depends on the weight associated with the edge connecting them.

The messages are exchanged in an iterative manner between variable and check nodes for a predefined number of iterations or until the decoding achieves convergence. The final LLR value of each variable node after $L$ iterations of the BP decoding is calculated as follows:
\begin{equation}
    m^{(L)}_{v}=\sum_{c\in \mathcal{N}_v} m_{c\rightarrow v}^{(L)}(w_v).
\end{equation}
These LLRs are then passed to the decoder of the precode to find the original $k$ information symbols.

\section{The Density Evolution Analysis of AFC}
In this section, we focus on the AFC code only without the precoder to better understand the AFC code performance and be able to optimise the code parameters. We will use the density evolution (DE) analysis \cite{richardson2001capacity} to evaluate the performance of the BP decoder of AFC. DE has been extensively applied to graph-based codes with static uniformity, which involves tracking the distribution of the messages exchanged along the edges of the bipartite graph in each iteration of the message passing algorithm (MPA) \cite{tanner1981recursive}. 

The DE analysis of AFC is different than that for the binary linear codes and in particular LDPC codes. In LDPC codes, or any binary linear block code, the information symbols are XORed according to the generator matrix to generate coded symbols. In the decoding of LDPC codes, a bipartite graph is generated based on the parity check matrix of the code, where the check operation simply checks whether the XOR of the variable nodes connected to the check node is 0 or not. The DE analysis of LDPC codes is then defined on this graph and the updating rules at variable and check nodes are defined according to these XOR operations \cite{chen2002density}. Although the encoding process of AFC can be represented by a bipartite graph, the operation at check nodes is different. As can be seen in Fig. \ref{fig:fig10}, to generate each AFC coded symbols, a set of randomly chosen information symbols, are first BPSK modulated, and then summed using the real weight coefficients (assigned to the edges in the bipartite graph). Furthermore, similar to the DE analysis of Raptor codes \cite{jayasooriya2017analysis}, the bipartite graph of AFC is constructed based on the generator matrix of the code. The weighted sum at the check nodes of AFC will make tracking the evolution of the messages in the BP decoder very challenging.

It is important to note that the DE analysis is valid for the asymptotically long block length regime, where the message and codeword lengths go to infinity. In Section V we will focus on the design of the precode for AFC to optimise the performance in the short block length regime. In fact, the optimised weight set obtained from DE in the asymptotic regime  can be used in the short block length regime with the help of a precode with a powerful error-correcting capability to achieve the desired level of reliability.

\subsection{The Density Evolution Analysis}
Without loss of generality, we assume that all incoming messages in a node are independent and identically distributed (i.i.d). This assumption was first considered in \cite{richardson2001capacity} and further justified in \cite{wang2005density}. Based on this assumption, the bipartite graph can be viewed as a set of independent sub-trees with independently distributed messages, making the analysis more feasible. With probability arbitrarily approaching to 1 when $n$ goes to infinity, a cycle-free bipartite graph emerged  \cite{wang2005density}. We would like to emphasise that the assumption is valid for asymptotically long block lengths only. For short block lengths, this assumption does not hold unless the bipartite graph is cycle-free. The DE analysis in this section is therefore valid for the infinite block length regime. Later in Section V, we will focus on the design of the precoder to optimise the AFC performance at short block lengths.

The other assumption which has been widely considered for the DE analysis, is the all-zero codeword transmission. For this condition to be met, the output channel LLRs should be symmetric; that is, the bit error rate is independent of the transmitted codeword \cite{dupraz2014density}. However, AFC does not meet this condition as the average power of the output symbols depends on the information sequence. For example, when the all-zero information sequence is being encoded using AFC, each coded symbol will be equal to $\sum_{i=1}^{d_c}w_i$, which is equivalent to the signal with the highest power assuming that $w_i$'s are all positive. For information sequences with some non-zero symbols, some of the coded symbols will have lower power. This results in unequal protection of information sequences. To solve this problem and meet the DE analysis requirements, we adopt the idea of the i.i.d channel adapter \cite{hou2003capacity}.

In particular, we slightly modify the encoding process of AFC in \eqref{eq:afcenc} by multiplying a binary random number, i.e., $+1$ and $-1$, with the weight associated with each edge in the bipartite graph. The $i$th AFC coded symbols are generated as follows:
\begin{align}
    c_i=\sum_{j=1}^{d}(-1)^{t_j}w_jv_{i,j},
    \label{eq:modifiedenc}
\end{align}
where $t_j$'s are uniformly and randomly drawn from set $\{0,1\}$. In the rest of the paper, we use this modified encoder unless otherwise specified.

\subsubsection{DE Check Node Updating Rule}
It is important to note that, unlike binary graph-based codes that are mainly characterised by the degree distribution function in the asymptotic block length regime, the AFC analysis should also take into account the weights associated with each edge in the graph. To do so, we assume that the messages passed from check to variable nodes are weight dependent. Let $m^{(\ell)}_{c\rightarrow v}(w)$ denote the message passed from check node $c$ to variable node $v$ along the edge with weight $w$ in the $\ell$th iteration of MPA. Let $f^{(\ell)}_{cv}(w,m)$ denote the $\mathrm{pdf}$ of $m^{(\ell)}_{c\rightarrow v}(w)$. In DE, we track the evolution of $f^{(\ell)}_{cv}(w,m)$. 

To calculate $f^{(\ell)}_{cv}(w,m)$, we need to find the $\mathrm{pdf}$ of $m^{(\ell)}_{c\rightarrow v}(w)$ which is derived in \eqref{eq:ctov}. Due to the complexity of this equation, it is not straightforward to find $f^{(\ell)}_{cv}(w,m)$. To address this, we propose to collect the exchanged messages by random sampling, i.e., by performing Monte-Carlo simulations. In particular, we randomly generate samples of $y$ assuming that an all zero-codeword is being sent. It is important to note that for DE we consider channel adapters and therefore the modified encoder in \eqref{eq:modifiedenc}. We also randomly generate samples of $p_{v\rightarrow c}^{(\ell-1)}(b_v)$ from the $\mathrm{pdf}$ of $m_{v\rightarrow c}^{(\ell-1)}$, denoted by $f^{(\ell-1)}_{vc}(m)$.

More specifically, to generate samples of $p_{v\rightarrow c}^{(\ell-1)}(b_v)$, we first draw a random number $m_{v\rightarrow c}^{(\ell-1)}$ from $f^{(\ell-1)}_{vc}(m)$. Then by using \eqref{eq:mtop},  $p_{v\rightarrow c}^{(\ell-1)}(b_v)$ is calculated for $b_v=1$ and $b_v=-1$. These will be inserted into \eqref{eq:ctov} to calculate one sample of $m^{(\ell)}_{c\rightarrow v}(w)$. Once a large number of samples are generated, $f^{(\ell)}_{cv}(w,m)$ can be approximated by finding the histogram of samples of $m^{(\ell)}_{c\rightarrow v}(w)$.

\subsubsection{DE Variable Node Updating Rule}
\begin{figure}[t]
    \centering
    \includegraphics[width=0.95\columnwidth]{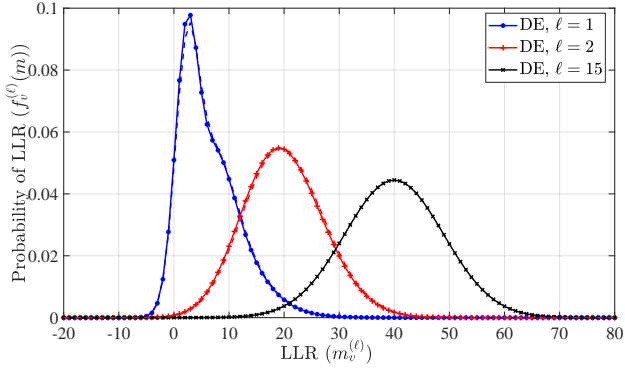}
    \caption{Variable node densities calculated by DE at different decoding iterations for an AFC code with $n=8000$, $\mathcal{W}=\{0.5097,0.4992,0.4960,0.4949\}$, $d_c=4$, and rate $R=0.5$, when $\gamma=10$dB. Solid and dashed curves respectively show DE analytical results and simulation results.}
    \label{fig:fig13}
\end{figure}

Let $f^{(\ell)}_{vc}(m)$ denote the $\mathrm{pdf}$ of the $m^{(\ell)}_{v\rightarrow c}$, which is given in \eqref{eq:vtoc}. As we assume that the messages passed along the edges are independent in an asymptotically long block length regime, we can find $f^{(\ell)}_{vc}(m)$ as follows:
\begin{align}
    f^{(\ell)}_{vc}(m)=\frac{1}{2d_c}\bigotimes_{i=1}^{d_v-1} \bigotimes_{w\in\mathcal{W}^{\pm}} f^{(\ell)}_{cv}(w,m),
    \label{eq:densityvc}
\end{align}
where $\mathcal{W}^{\pm}=\left\{w,-w:~w\in\mathcal{W}\right\}$ is the set of all positive and negative weight coefficients due to channel adapter. This equation follows from the fact that each variable node has $d_v$ connected check nodes and the edges have weights which are randomly drawn from $\mathcal{W}^{\pm}$. Since each edge can randomly choose one of the $2d_c$ available weight coefficients, we normalise the density by multiplying it with $\frac{1}{2d_c}$. Fig. \ref{fig:fig13} shows the densities $m^{(\ell)}_{v}$ at different iterations of the BP decoding for an AFC code with weight set {$\mathcal{W}= \{0.5097,0.4992,0.4960,0.4949\}$}, degree $d_c=4$, and rate $R=0.5$, when SNR$=10$dB. As can be seen in this figure, the densities are shifting towards right when the iteration number increases.

\subsubsection{Approximation of the bit error rate using DE}
\begin{figure}[t]
    \centering
    \includegraphics[width=0.95\columnwidth]{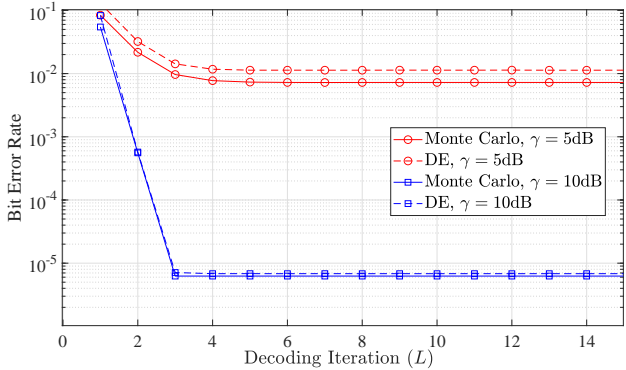}
    \caption{The BER of AFC at different decoding iterations, when $n=8000$, $d_c=4$, $\mathcal{W}=\{0.8632, 0.4495, 0.2300, 0.0004831\}$, and rate $R=0.5$}.
    \label{fig:figDE}
\end{figure}

We assume that DE converges after a few iterations and the check to variable node densities converge to $f^{(\infty)}_{cv}(w,m)$. The variable node density  denoted by $f^{(\infty)}_v(m)$ is then given by:
\begin{align}
    f^{(\infty)}_{v}(m)=\frac{1}{2d_c}\bigotimes_{i=1}^{d_v} \bigotimes_{w\in\mathcal{W}^{\pm}} f^{(\infty)}_{cv}(w,m),
    \label{eq:densityv}
\end{align}
The bit error rate (BER) of the AFC code, denoted by $\epsilon\left(d_c,d_v,\mathcal{W}\right)$, can then be calculated as follows:
\begin{align}
\epsilon\left(d_c,d_v,\mathcal{W}\right)=\int_{-\infty}^0 f^{(\infty)}_{v}(x)dx,
    \label{eq:berde}
\end{align}
which directly follows from the assumption of all-zero information sequence and that a bit error occurs when the LLR is calculated to be negative. Fig. \ref{fig:figDE} shows the approximation of BER at different decoding iterations using the DE analysis for an AFC code at different SNRs when $n=8000$. As can be seen, \eqref{eq:berde} provides a tight approximation for the BER for AFC codes when the information block length $n$ is large.

We note that due to the weighted sum of (bipolar) information symbols when generating AFC coded symbols, the $\mathrm{pdf}$ of the message sent from check to variable nodes cannot be represented in closed form. Instead, we used a sampling approach to find the $\mathrm{pdf}$ of the message sent form check to variable nodes, which can be easily implemented to numerically find the BER of AFC codes using \eqref{eq:berde}.

\section{AFC Weight Set Optimisation}
In this section, we use the DE analysis and define an optimisation problem to minimise the BER of the AFC and find the optimal weight set for a given check node degree $d_c$, variable node degree $d_v$, and channel SNR. The optimisation problem can be summarised as follows:
\begin{align}
\label{eq:opt}
    \min_{\mathcal{W}}&~~\epsilon(d_c,d_v, \mathcal{W})\\
    \nonumber\mathrm{s.t.}~~~~~&\\
    \nonumber\mathrm{C_1:}&~w_i>0,~\text{for}~i=1,\cdots,d_c,\\
    \nonumber\mathrm{C_2:}&~\sum_{i=1}^{d_c}w_i^2\approx 1,
\end{align}
where $\epsilon(d_c,d_v, \mathcal{W})$ is found via \eqref{eq:berde}, condition $\mathrm{C_1}$ is to make sure that all weight coefficients are positive, and condition $\mathrm{C_2}$ it to make sure the average power of the AFC coded symbols is 1. Traditionally, the optimisation of graph-based channel codes involved maximising the decoding threshold. This is equivalent to minimising the bit error rate, which we consider in this paper for AFC. We further note that we are interested in finding the optimal weight set for a given fixed code degree $d_c$. A more general optimisation can be easily defined to determine the optimal degree distribution function and weight set jointly. This is, however, out of the scope of this paper.

The output of DE analysis is usually the BER or the threshold of the code in terms of noise power. To be able to characterise the messages and run DE we need to set the SNR and the code rate. The methodology used is similar to the approach used for optimising the rateless codes, and in particular Raptor codes \cite{shokrollahi2006raptor}. That is the code is analysed at a fixed rate and a given SNR and the degree distribution is accordingly optimised \cite{richardson2001capacity,jayasooriya2017analysis,8374935,8052467} to minimise the BER. Our objective function in \eqref{eq:opt} is to minimise the bit error rate at a given SNR and code rate, i.e., given $d_c$ and $d_v$. We will show in Section V that the optimised weight set performs reasonably well across all SNRs when a proper precoder is used.

We use the differential evolution algorithm \cite{storn1997differential} to solve \eqref{eq:opt}. The differential evolution is a simple yet powerful optimisation tool based on the population stochastic search technique. There are three main parameters that control the optimisation algorithm: the scaling factor, crossover probability, and population size. The setting of these three parameters would directly affect the time and performance of the optimisation process. The population of differential evolution may move through a different region of search space to find suitable candidates. Although this approach is time-consuming, it suffices our purpose as we do not aim for efficient optimisation and instead would like to find some good weight sets for AFC. 

For the case of AFC, the optimisation process involved the tuning of weight-set with the prefixed rate, $R_{\mathrm{AFC}}=d_c/d_v$, at the given SNR, $\gamma$, to minimise the BER as in \eqref{eq:opt}. Table \ref{table:Table1} shows the optimised AFC weight set for different degree $d_c$, when the rate is $R_{\mathrm{AFC}}=2$ and $\gamma=15$dB. In this optimisation, we performed 100 iterations of DE and generated 1000 samples of LLRs to calculated the densities in \eqref{eq:densityvc}. As for the differential evolution, we considered the population size of approximately 50, crossover probability of 1, and mutation factor of 0.85. 
\begin{table}[t]
\centering
\caption{AFC Optimised weight sets obtained by \eqref{eq:opt}, when $\gamma=15$dB and $R_{\mathrm{AFC}}=2$.}
     \label{table:Table1}
     \scriptsize
 \begin{tabular}{|c|c|c|} 
 \hline
  &  $d_c$ & Weight Set\\ 
 \hline
$\mathcal{W}_1$ &2 & \{0.7202, 0.6938\}\\
\hline
$\mathcal{W}_2$  & 3 & \{0.7050, 0.5234, 0.4786\} \\
 \hline
$\mathcal{W}_3$  & 4 &\{0.8632, 0.4495, 0.2300,  0.0004831\}\\
 \hline
$\mathcal{W}_4$  & 5 &\{0.7272, 0.5014, 0.3151, 0.2921, 0.0754\}\\
 \hline
 $\mathcal{W}_5$  & 6 &\{0.8006,0.4914, 0.2357, 0.1802, 0.1713, 0.0174\}\\
 \hline
  $\mathcal{W}_6$  & 7 &\{0.7846, 0.4197, 0.4023, 0.1522, 0.1151, 0.0739, 0.0676\}\\
 \hline
 \end{tabular}
\end{table}

\begin{table}[t]
\centering
\caption{Benchmark weight sets in the literature for AFC when $d_c=4$.}
\label{table:2}
\scriptsize
  \begin{tabular}{|c|c|} 
 \hline
  Name & Weight Set\\ 
 \hline
$\bar{\mathcal{V}}_1$ \cite{shirvanimoghaddam2013near}& \{0.9103, 0.3641, 0.1655, 0.1071\} \\
\hline
$\bar{\mathcal{V}}_2$ \cite{shirvanimoghaddam2013near} & \{0.8902, 0.3815, 0.2054, 0.1406\}\\
\hline
$\bar{\mathcal{V}}_3$ \cite{zhang2020finite} & \{0.7303, 0.5477, 0.3651, 0.1826\}\\
\hline
$\bar{\mathcal{V}}_4$ \cite{cui2011seamless} & \{0.6576, 0.6576, 0.3288, 0.1644\}\\
\hline
$\bar{\mathcal{V}}_5$ \cite{abbas2019performance} & \{0.8686, 0.4329, 0.2159, 0.1075\}\\
\hline
$\bar{\mathcal{V}}_6$ \cite{rao2019low} & \{0.6325, 0.6576, 0.3162, 0.1644\}\\
\hline
 \end{tabular}
\end{table}

When optimising the weight set of the AFC using (\ref{eq:opt}), we need to specify the rate and SNR. In other words, the weight set obtained through (\ref{eq:opt}) depend on the rate and channel SNR. For example, when $R_{\mathrm{AFC}}=0.5$ and $\gamma=5$dB, the optimised weight set we obtained for $d_c=4$  is $\mathcal{W}_3^{*}=\{0.5097,0.4992,0.4960,0.4949\}$. When $R_{\mathrm{AFC}}=2$ and $\gamma=15$dB, the optimised weight set we obtained for $d_c=4$  will be $\mathcal{W}_3=\{0.8632, 0.4495, 0.2300, 0.0004831\}$. The optimised weight coefficients seem to be closer to each other when optimising AFC at lower SNRs and rates. 

\begin{figure}[!t]
    \centering
    \includegraphics[width=0.95\columnwidth]{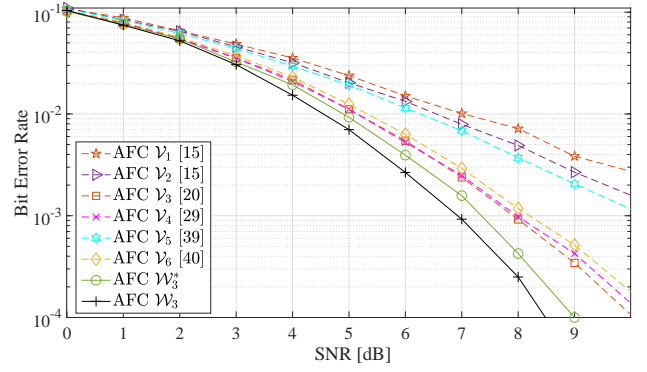}
    \caption{The BER of an  AFC with $n=8000$, $R_{\mathrm{AFC}}=0.5$, and optimised weight set $\mathcal{W}_3$ (Table \ref{table:Table1}), in comparison to previously designed weight sets in the literature (Table \ref{table:2}). The weight set $\mathcal{W}_3^{*}=\{0.5097,0.4992,0.4960,0.4949\}$ was optimised for AFC at $R_{\mathrm{AFC}}=0.5$ and $\gamma=5$dB.}
    \label{fig:figLongAFCOpt}
\end{figure}

Fig. \ref{fig:figLongAFCOpt} compares the BER performance of AFC with these two weight sets, $\mathcal{W}_3$ and $\mathcal{W}^{*}_3$, when $R_{\mathrm{AFC}}=0.5$. As can be seen, the weight set optimised at a high SNR and higher rate performs better compared to weight set optimised at a low SNR and low rate. This observation provides us a guideline on the input parameter selection for the optimisation process, i.e., the rate and SNR should be as high as possible within the region of investigation to obtain better performance. We also compared the optimised weight sets with some weight sets previously used in the literature (see Table \ref{table:2}) \cite{shirvanimoghaddam2013near,zhang2020finite,cui2011seamless,abbas2019performance,rao2019low}. Fig. \ref{fig:figLongAFCOpt} shows the superiority of the optimised weight sets for AFC in terms of BER. 

\section{Design of AFC for Short Packet Communications}

In this section, we evaluate the precoded AFC in the short block length regime. We focus on short messages, from only a few ($\le$100) bits up to a few hundred ($\le$1000) bits. This has been widely considered in the literature and standardisation documentations for 5G URLLC. Please note that DE analysis is not applicable in the short blocklength regime. Thus, we instead use the optimised parameter obtained from DE in the long block length, and then further optimise the code performance by tuning the precoder.

In our work, we selected BCH codes and LDPC codes as precoder of AFC. There are two criteria that we consider when select a precoder for AFC: 1) error correction capabilities and 2) decoding complexity. BCH codes are chosen as precoder of AFC for very short message length as it is a very powerful error correcting codes. Although LDPC codes benefits from their low complexity message passing decoder, they do not perform well in the short block length regime. LDPC codes, however, show good performance with low decoding complexity at moderate and high message length. It is worth to note that the proposed framework is compatible with other fixed-rate code. In fact any fixed rate code can be used as a precoder of AFC. The choice of the precoder must be aligned with the one of the criteria aforementioned, if not fulfill both of the criteria. In this case, BCH is chosen for very short message because it has almost the highest minimum Hamming distance and that can be efficiently decoded by OSD decoders. For moderate to long message lengths, we chose LDPC mainly because of their low complexity decoder.  

We use the normal approximation \cite{polyanskiy2010channel} as a benchmark for comparing the performance of AFC at short block lengths. For the AWGN channel, the normal approximation for the achievable rate is given by  \cite{polyanskiy2010channel}:
\begin{equation}
R\approx C-\sqrt{\frac{V}{n}}Q^{-1}(\epsilon)+\frac{\log_2n}{2n},
\end{equation}
where $C=\frac{1}{2}\log_2(1+\gamma)$ is the channel capacity, $\gamma$ is the channel SNR, $V=\log^2_2(e)\frac{\gamma(\gamma+2)}{2(\gamma+1)^2}$ is the channel dispersion, $\epsilon$ is the block error rate, and $Q(x)=\frac{1}{\sqrt{2\pi}}\int_{x}^{\infty}e^{-\frac{x^2}{2}}dx$ is the standard $Q$-function.

\subsection{Block Error Rate of AFC at Fixed Rate}
We first investigate the block error rate (BLER) performance of precoded AFC truncated at a fixed block length.  In particular, we treat AFC as a fixed-rate code in order to investigate the achievable reliability guarantees. This is essential to understand the effect of the precode rate on the overall performance of AFC codes.

\begin{figure}[!t]
    \centering
    \includegraphics[width=0.95\columnwidth]{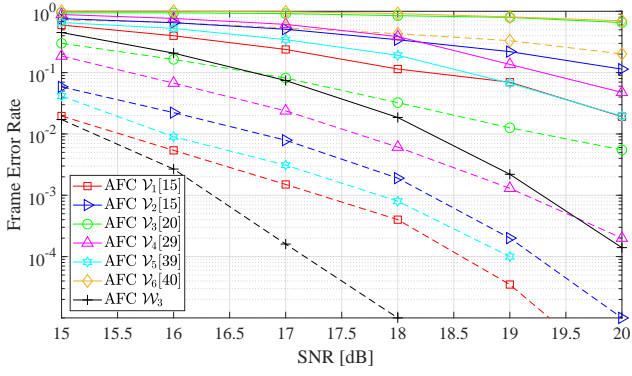}
    \caption{The BLER of a precoded AFC using weight $\mathcal{W}_3$  (Table \ref{table:Table1}), in comparison to previously designed weight sets  (Table \ref{table:2}). Solid line show the results for the case with $R_{\mathrm{AFC}}=4.46$ and $\mathcal{R}=2$. Dashed lines show the results for the case with $R_{\mathrm{AFC}}=3.34$ and $\mathcal{R}=1.5$. A BCH(127,57) is used as the precoder.}
    \label{fig:figBLERHighSNR1}
\end{figure}

Fig. \ref{fig:figBLERHighSNR1} shows the performance of the precoded AFC when a BCH(127,57) code is used as a precoder. As can be seen in this figure, the optimised weight set ($\mathcal{W}_3$ in Table \ref{table:Table1}) outperforms other weight sets previously designed for AFC. The improvement in terms of BLER is consistent when the AFC is operating at different overall rates. When BCH(63,57) code is used as a precoder, results are consistent with what we observed in Fig. \ref{fig:figBLERHighSNR1} and that the optimised weight set obtained in this paper outperforms weight sets previously designed for AFC.

\begin{figure}[!t]
    \centering
    \includegraphics[width=0.95\columnwidth]{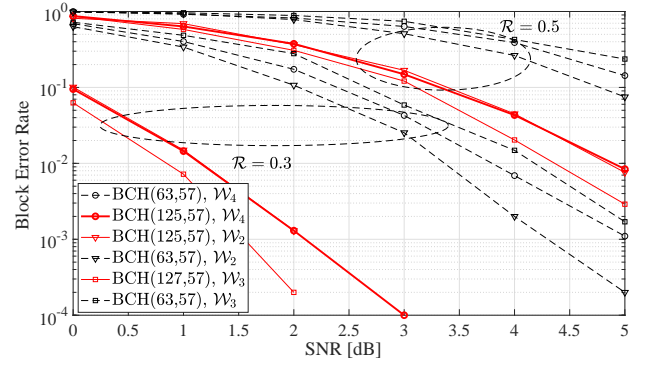}
    \caption{BLER versus SNR for precoded AFC, when  BCH(63,57) and BCH(127,57) are used as precoder, at low SNRs with optimised weight $\mathcal{W}_3$ (Table \ref{table:Table1}).}
    \label{fig:Low_SNR_W234}
\end{figure}

Fig. \ref{fig:Low_SNR_W234} shows the comparison between AFC codes precoded with different BCH codes, i.e., BCH(63,57) and BCH(127,57), where the message length is $k=57$, and at different overall  rates. As can be seen, the AFC code with a low-rate precoder achieves a lower BLER. This is mainly because the lower rate precoder is more capable of correcting residual errors that AFC is unable to recover, especially when the SNR is low. The performances of BLER at high SNRs shows the same trend, where the lower rate precoder performs better in terms of BLER at different rates.

Fig. \ref{fig:Low_SNR_W234} also shows the performance of precoded AFC using different weight sets. It can be observed that the lower rate precode performs better than the high rate precode when using different weight sets with different degrees. It is important to note that when a high rate precode, i.e., BCH(63,57) is being used, the weight set $\mathcal{W}_2$ with degree $d_c=3$ outperforms the other weight sets. However, when the higher rate precode, i.e., BCH(127,57) is being used, the weight set $\mathcal{W}_3$ with degree $d_c=4$ outperforms the other weight sets in different overall rates. This shows that the degree of the AFC code should be chosen carefully, and that depends on the precode rate in order to minimise the BLER. 

\begin{figure}[!t]
    \centering
    \includegraphics[width=0.95\columnwidth]{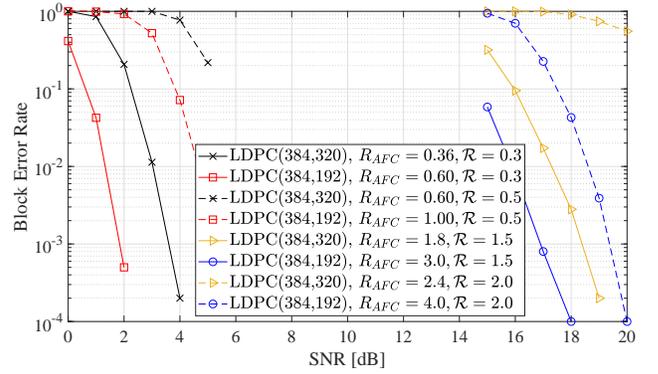}
    \caption{BLER versus SNR for precoded AFC when  LDPC(384,320) and LDPC(384,192) are used as precoder at low SNRs and high SNRs with weight $\mathcal{W}_3$ (Table \ref{table:Table1}).}
    \label{fig:figBLERLdpcLowHighSNR}
\end{figure}

Simulation results demonstrate that a similar trend is obtained when using a LDPC code as the precoder, as shown in Fig. \ref{fig:figBLERLdpcLowHighSNR}. That is AFC with a low rate precoder outperformed AFC with the high rate precoder in terms of BLER.

\subsection{Achievable Realised Rate}

We compare the realised rates achievable by the optimised weights versus previously designed weight sets in Table \ref{table:Table1}. For all simulations, we assumed that the receiver attempts decoding every time it receives $\delta=5$ additional AFC coded symbols. The first decoding attempts occurs when the receiver collects $m_0=\frac{2k}{\log(1+\gamma)}$ AFC coded symbols. The decoder sends an acknowledgment to the transmitter when the decoding succeeds and accordingly the transmission is terminated. 

\begin{figure}[t]
    \centering
    \includegraphics[width=0.95\columnwidth]{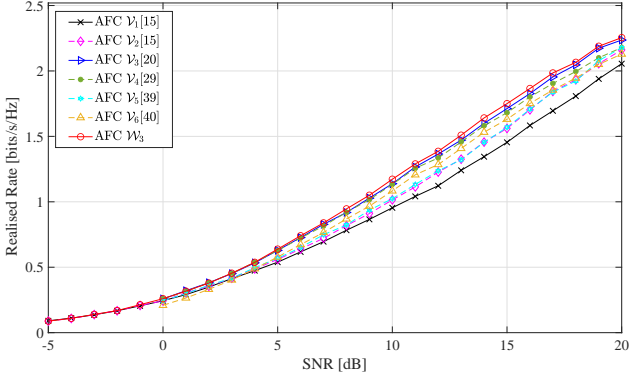}
    \caption{The realised rate of precoded AFC using our optimised weight set $\mathcal{W}_3$  (Table \ref{table:Table1}), in comparison to previously designed weight sets in the literature (Table \ref{table:2}). The precoder used here is the  BCH(63,57) code.}
    \label{fig:figRateBCHd3}
\end{figure}

Results are shown in Fig. \ref{fig:figRateBCHd3}. It is clear that our optimised weight set $\mathcal{W}_3$ is superior in performance to previous AFC weight set designs, with similar degrees, i.e., similar complexity. In particular, for an AFC code precoded with  BCH(63,57), our optimised weight set $\mathcal{W}_3$ can achieve about 4.35\% and 9.68\% higher realised rate than weight sets $\mathcal{V}_1$ and $\mathcal{V}_2$, respectively, in the high SNR regime (around 20dB), and 13.96\% and 18.49\% higher realised rates, respectively,  in the low SNR regime (around 5dB). When compared with benchmark weight sets $\mathcal{V}_3$ and $\mathcal{V}_4$ (Table \ref{table:2}), the performances of the benchmark weight sets are quite close with our optimised weight set $\mathcal{W}_3$, but our optimised weight set still demonstrated superiority compared to benchmark weight sets. Our optimised weight set $\mathcal{W}_3$ can achieve about 0.84\% and 3.37\% higher realised rate than weight set $\mathcal{V}_3$ and $\mathcal{V}_4$, respectively, in the high SNR regime (around 20dB), and 1.49\% and 1.66\% higher realised rates, respectively, in the low SNR regime (around 5dB).

\begin{figure}[!t]
    \centering
    \includegraphics[width=0.95\columnwidth]{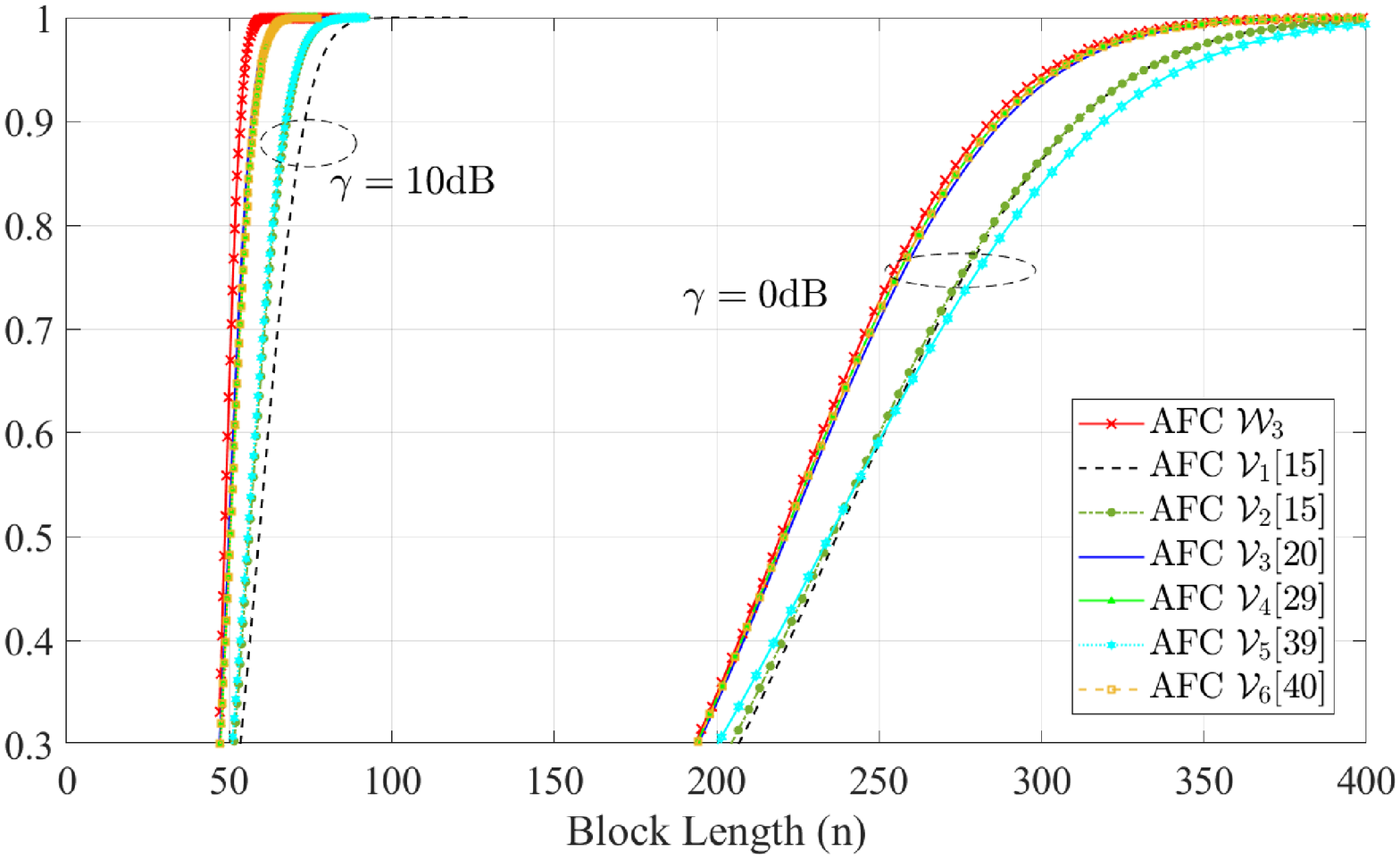}
    \caption{The $\mathrm{cdf}$ of the block length for a precoded AFC using our optimised weight set $\mathcal{W}_3$ (Table \ref{table:Table1}), in comparison to previously designed weight sets in the literature (Table \ref{table:2}). The precoder used here is the  BCH(63,57) code.}
    \label{fig:figCDFBCHd3}
\end{figure}

It is important to note that, in theory, the realised rate is defined for the zero-error transmission. However, due to computational limitations, such a realised rate cannot be calculated in practice. Thus, here we assume that the plotted realised rates correspond to block error rates less than $10^{-4}$, i.e., for $10^4$ messages, no errors are exhibited.  Furthermore, since the transmitted block length varies from one frame to the next, we also plot the cumulative distribution function ($\mathrm{cdf}$) of the block length at different SNRs to understand better what latency guarantees AFC can provide. As shown in Fig. \ref{fig:figCDFBCHd3}, our optimised code exhibits a smaller variance in the block length than previously designed AFC codes. Thus, our optimised weight set can provide better latency guarantees, particularly for delay-sensitive applications with little tolerance for jitters.

\begin{figure}[t]
    \centering
    \includegraphics[width=0.95\columnwidth]{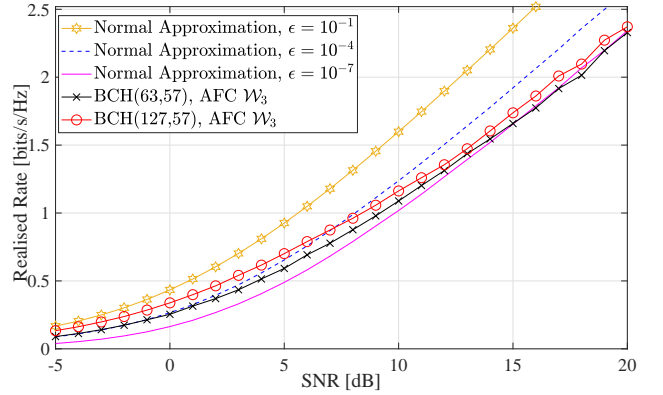}
    \caption{The realised rate of an AFC precoded with a BCH code using weight set $\mathcal{W}_3$ (Table \ref{table:Table1}).}
    \label{fig:figRateBCHd3PPV}
\end{figure}

\begin{figure}[!t]
    \centering
    \includegraphics[width=0.95\columnwidth]{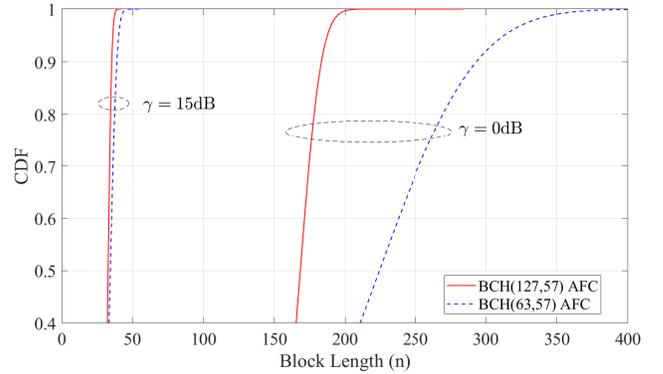}
    \caption{The cumulative distribution function ($\mathrm{cdf}$) of the block length of a precoded AFC with weight set $\mathcal{W}_3$ (Table \ref{table:Table1}) and different precoder rates.}
    \label{fig:figCDFBCHd3PPV}
\end{figure}

Results are shown in Fig. \ref{fig:figRateBCHd3PPV}, where we  consider two precoders, BCH(63,57) code and BCH(127,57). We choose these two precoders, such that they have the same message length but different rates. For AFC, we use the optimised weight set $\mathcal{W}_3$ from Table \ref{table:Table1}. As can be seen, the realised rate of AFC with a lower precoder rate  is higher than an AFC with a higher rate precoder over a wide range of SNRs. At $\gamma=20$dB, AFC with BCH(127,57) has a gap of 9.57\% to the PPV bound compared to AFC with BCH(63,57), which has a gap of 13.68\% to the bound. At $\gamma=5$dB, AFC with BCH(127,57) is closer to bound, i.e., it has a gap of 7.14\% to the bound compared to AFC with BCH(63,57) that has a gap of 9.57\% to the bound.  Similarly, in Fig. \ref{fig:figCDFBCHd3PPV}, the $\mathrm{cdf}$ of the block length has a smaller variance, i.e., steeper gradient, when the precoder rate is lower. A similar observation was made in \cite{jayasooriya2017analysis} for the case of LDPC precoded Raptor codes. 


\begin{figure}[!t]
    \centering
    \includegraphics[width=0.95\columnwidth]{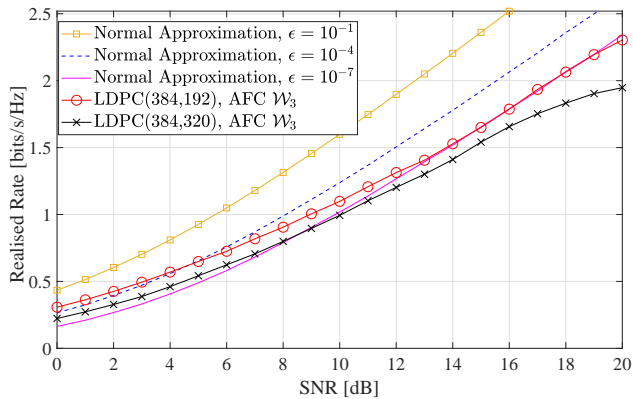}
    \caption{The realised rates of a precoded AFC using weight set $\mathcal{W}_3$ (Table \ref{table:Table1}) and different LDPC precoder rates.}
    \label{fig:figLDPC}
\end{figure}

The observation that lower precoder rates have the potential to achieve higher realised rates over a wide range of SNRs is further validated in Fig. \ref{fig:figLDPC}. In Fig. \ref{fig:figLDPC}, we use a range of LDPC precoders with different rates and longer block lengths. Results for these medium length blocks show the same trend as those of short blocks in Fig. \ref{fig:figRateBCHd3PPV}. When LDPC code is used as the precoder of AFC, the $\mathrm{cdf}$ of block length of the respective AFC with different precoder rates at two regions of SNRs show similar trend with what we observed Fig. \ref{fig:figCDFBCHd3PPV} using BCH codes as precoder. That by using a lower precoder rate for AFC, the $\mathrm{cdf}$ of the block length has a steeper gradient compared to an AFC with a higher rate precoder. This result further strengthen our claim that lower precoder rate has superior performance compared to higher precoder rate. We do not show the respective resuls for LDPC precoded AFC due to the space limitation.  

It is well established that the degree of the AFC code plays a significant role in its achievable rates, i.e., there is a trade-off between the maximum achievable rate and the allowed encoding/decoding complexity. More specifically, in a noise-free environment, the AFC code can achieve a rate of $d_c$. However, its decoding complexity increases exponentially with $d_c$ as can be clearly inferred from Section VI. 

\begin{figure}[!t]
    \centering
    \includegraphics[width=0.91\columnwidth]{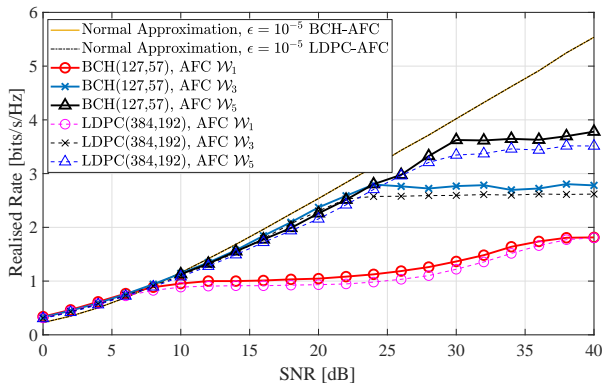}
    \caption{Realised rate comparison for different degree of optimised weight: $\mathcal{W}_1$, $\mathcal{W}_2$ and $\mathcal{W}_3$ (Table \ref{table:Table1}) with BCH (127,57) and LDPC (384,192) as a precoder.} 
    \label{fig:BCHandLDPCDifferentDegree}
\end{figure}

We expect that the check node degree to be the upper bound on the maximum achievable rate when the SNR is sufficiently high. When the BCH(127,57) code is used as the precoder, the achievable rates are plotted in Fig. \ref{fig:BCHandLDPCDifferentDegree}. On the same graph, the achievable rates for precoded AFC with LDPC(384,192) as precode are plotted. We consider three different weight sets, $\mathcal{W}_1$, $\mathcal{W}_3$, and $\mathcal{W}_5$ from Table \ref{table:Table1}, for AFC code with degree $d_c=2$, $d_c=4$, and $d_c=6$, respectively. As can be seen in Fig. \ref{fig:BCHandLDPCDifferentDegree}, the AFC code with the larger degree achieves a higher realised rate over a wide range of SNRs. 

In this work, we only provided the comparisons between our optimised precoded AFC and other designs of AFC available in the literature, instead of comparing with the other state-of-the-art codes, i.e., LDPC codes, Polar codes, and RM codes. This is because these codes are fixed rate codes. They are mainly designed for particular SNRs and at desired code rates. As for other binary rateless codes, such as Raptor codes, they are designed for the binary field and thus they should be equipped with higher order modulation to achieve higher spectrum efficiency when operating over the wireless channel. Thus, it is not fair to compare with our optimised precoded AFC, which is designed for rateless transmission over wireless channel where channel state information is not known at the transmitter. 

\subsection{Realised Rate over the Fading Channel}
We now consider a fading channel, which is commonly present in vehicular communications. Since AFC is a rateless code, the encoder can send as many coded symbols are required by the receiver to perform a successful decoding. This means that the transmitter can adapt to the channel condition without knowing the channel condition in advance. 

\begin{figure}[!t]
    \centering
    \includegraphics[width=0.95\columnwidth]{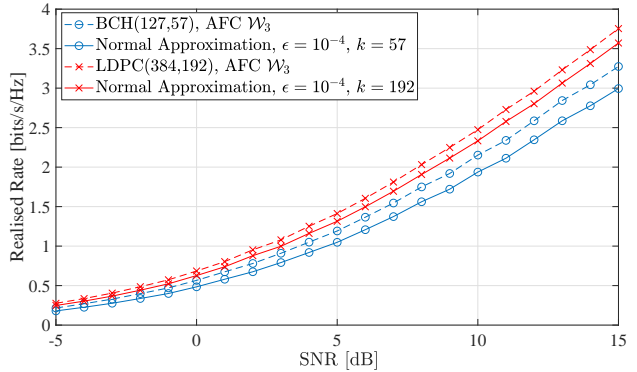}
    \caption{The realised rates of a precoded AFC with LDPC(384,192)  and BCH(127,57) as precoder and $\mathcal{W}_3$ (Table \ref{table:Table1}) as the weight set over a quasi-static Rayleigh fading channel.}
    \label{fig:Fading}
\end{figure}

We assume a quasi-static Rayleigh fading channel, where the channel remains fixed over the transmission of a message block and varied independently between blocks. Fig. \ref{fig:Fading} shows the realised rate of the precode AFC, when $\mathcal{W}_3$ from Table \ref{table:Table1} is used as the weight set. As can be seen in this figure, AFC can closely approach the finite block length bound across a wide range of channel conditions. This is mainly because each instance of the fading channel can be realised as an AWGN channel, and AFC already performs close to the finite length bound, without channel knowledge at the transmitter side. This also suggests that the AFC will perform well under more complex fading scenarios with correlated fading. Therefore, the code designed over the AWGN channel can be effectively used for the fading scenario. As for multi-path and fast fading scenario, we note that other approaches, like OFDM and interleaving, are required which are beyond the scope of this work. In this paper, we mainly focused on the design of AFC codes and tried to provide a thorough discussion on the effect of code parameters on the performance.

\section{Complexity Analysis of AFC Decoding}
In this section, we provide a detailed analysis of the complexity of the encoder and decoder of the precoded AFC. The complexity of the encoder is mainly depend on the degree of AFC, $d_c$, as well as the number of coded symbols, $m$. The encoding complexity scale linearly with the block length $m$, since the generation of each AFC coded symbols is independent of other symbols, and that involves the weighted sum of $d_c$ randomly selected information symbols.

\subsection{The complexity of the BP decoder}
The BP decoder operates in an iterative manner, and each iteration involves exchanging messages between variable and check nodes. In particular, during variable node updates, $nd_v$ messages will be calculated and sent to check nodes. The variable node update operation \eqref{eq:vtoc} is simple and only involves the summation of $d_v-1$ incoming messages. Assuming that each summation requires 1 floating point operation (FLOP), the complexity of the entire variable node update round in approximately $\mathcal{O}(nd_v^2)$ FLOPs.

The check node update rule \eqref{eq:ctov} is more complex and involves calculating believes \eqref{eq:mtop}. This is mainly because of the weighted edges in the bipartite graph and the real sum operation at check nodes. In each check node update round, $md_c$  messages will be calculated and sent to variable nodes. Each calculation involves approximately $(d_c+2)\times 2^{d_c+1}$ FLOPs, assuming that the probabilities $p_v$ are exchanged between the variable and check nodes. Therefore, the complexity of the entire check node update round is approximately  $\mathcal{O}(m d_c^22^{d_c+1})$ FLOPs. Since we have $md_c=nd_v$ and $R=n/m$, the overall complexity per information symbol after $L$ iterations of the BP decoder is approximately given by $\mathcal{O}(L\frac{d_c^2}{R^2}(1+R2^{d_c+1}))$. It can be seen that the complexity of the decoder increases exponentially with the check node degree, which is common for all message passing decoders. 

Since AFC is using a similar decoder as rate compatible modulations (RCM) \cite{cui2011seamless} (in fact rate compatible modulation is a special case of AFC), several approaches proposed in \cite{lu2018low} can be used to reduce the complexity of the BP decoder. A common approach is to implement the convolution process in \eqref{eq:ctov} using Fast Fourier Transform (FFT). Another approach is to use the time-domain convolution with the application of Zigzag deconvolution structure or as known as tree-based convolution \cite{lu2018low}. LLR-based decoder is also another method proposed to reduce the decoding complexity \cite{rao2013log}. Using LLR offers implementation advantages over using probabilities, because multiplications are replaced by additions and the normalization step is eliminated. Several other approached have been proposed in \cite{lu2017parallel,rao2018efficient,rao2013log}, which can be applied to AFC codes. A comparison between different decoding approaches for AFC is available in \cite{lu2018low}. 

\subsection{Threshold-based Decoder for BCH precoded AFC} 
The precoded AFC requires two decoders, i.e.,  the BP decoder for AFC and the decoder for precode. The decoding process is carried out in a way that every time that a new set of AFC coded symbols are received, the decoder needs to run both BP and the decoder for precode. This process will stop only when the decoding succeeds. This leads to a huge complexity at the receiver side and accordingly the decoding time increases dramatically. When LDPC codes are used as precode, the complexity of the receiver can be minimised by jointly decoding AFC and LDPC codes using the BP decoder. The complexity of the LDPC decoder is negligible to the AFC decoder since all operations can be performed in the Log-domain and efficient algorithms have been already proposed. The problem becomes challenging when an OSD algorithm is being used for decoding the precode. When a low rate BCH code is used as the precode, we usually need a high-order OSD to achieve a near maximum-likelihood (ML) decoding performance. The complexity, however, increases with $\mathcal{O}(k^\ell)$, where $k$ is the message length and $\ell$ is the order of OSD. For a code with minimum Hamming distance $d_{\mathrm{H}}$, order $\ell=\lceil d_{\mathrm{H}}/4+1\rceil$ is asymptotically optimal, i.e., it can achieve near-ML performance \cite{Fossorier1995}. 

To reduce the decoding complexity, we propose a threshold-based decoding algorithm for the precoded AFC code. In this modified algorithm, we pass the soft information to the precode decoder and perform the algorithm only when the average reliability of the soft information is above a predefined threshold value. Further details of this algorithm can be found in Algorithm 1. Details on how to find the threshold value can be found in \cite{lim2019design}. It is important to note that in Step 9 of Algorithm 1, we check CRCs to verify whether the decoding succeeds or not. This is because the output of the OSD is always a valid codeword. Therefore, we need to add CRC bits to verify whether the output of the OSD is the transmitted codeword or not.

\begin{algorithm}[t]
\caption{Threshold-based Decoding for BCH}
\SetAlgoLined
\DontPrintSemicolon
Inputs: $\mathbf{y}$, $G_{\mathrm{pre}}$, $\delta$, $m_0$, $\gamma_{\mathrm{th}}$\;
initialisation: $m=m_0$\;
\While{(CRC check fails)}{
Request for additional $\delta$ AFC coded symbols\;
$m=m+\delta$\;
Perform BP decoding using $m$ AFC coded symbols\;
Calculate the average LLR of the output of the BP decoder, $\mathbb{E}[|L|]$\; 
\eIf{$\mathbb{E}[|L|]\ge\gamma_{\mathrm{th}}$}{
Perform OSD using $G_{\mathrm{pre}}$ and $\mathbf{L}$ and check CRCs
}{
Go to step 4\;
}
}
return $\hat{\mathbf{b}}$
\end{algorithm}

\begin{figure}[!t]
    \centering
    \includegraphics[width=0.95\columnwidth]{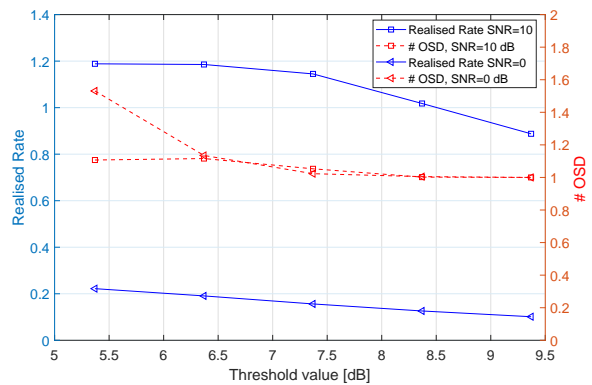}
    \caption{Number of OSD instances versus the threshold SNR for the BCH-AFC code at different SNRs when BCH(63,57) is used as a precoder.}
    \label{fig:figOSDRuntime}
\end{figure}
We evaluate the number of times that we need to run the OSD algorithm versus the threshold SNR. Results are shown in Fig. \ref{fig:figOSDRuntime}, where the average number of OSD instances and the realised rate are plotted versus the threshold SNR for a precoded AFC at different channel SNRs when BCH(63,57) was used as a precoder. As shown in this figure, the number of times that we run OSD is not very sensitive to the threshold value in high SNRs. However, the lower the threshold value, the higher the realised rate. At low SNRs, the threshold value cannot be chosen very small as it significantly increases the number of OSD instances. In fact, one can choose a very low threshold value and get a very high realised rate, but on the other hand, the complexity would significantly increase. It is important to note that if the threshold SNR is chosen very low, every time the receiver receives a new set of AFC symbols, it will run the OSD algorithm, no matter the reliability of AFC decoder outputs is. The proposed decoder can effectively limit the number of OSD instances to almost 1, when a proper threshold value is considered, with a negligible degradation in the realised rate \cite{lim2019design}. It is important to note that other approaches can be applied to further reduce the complexity of OSD algorithm as detailed in \cite{yue2021revisit}. This is, however, out of the scope of this work.

\section{Conclusion}
In this paper, we proposed a density evolution (DE) analysis framework for analog fountain codes. In the DE framework, we tracked the evolution of messages exchanged between the variable and check nodes when decoding an AFC using the belief propagation decoder. Using the proposed framework, we defined an optimisation problem to find the weight set of AFC. Results show that for the asymptotically long message lengths, the optimised weight sets outperform existing weight sets previously designed for AFC. We also studied the design of the precoder for AFC to optimise the performance at short block lengths. We mainly focused on BCH and LDPC codes and showed via simulations that a lower rate precoder could achieve a lower block error rate under the same overall rate and a higher realised rate over a wide range of signal to noise ratios (SNRs). We further showed that our optimised weight sets outperform existing weight sets in the literature at both low and high SNRs in the short block length regime. We further shed lights on how to reduce the decoding complexity of precoded AFC in the rateless setting. The proposed code can be effectively used for rateless transmission of short information sequences, which have applications in many mMTC and URLLC scenarios.

\bibliographystyle{IEEEtran}
\bibliography{IEEEabrv,Bibliography}

\begin{IEEEbiography}[{\includegraphics[width=1in,height=1.25in,clip,keepaspectratio]{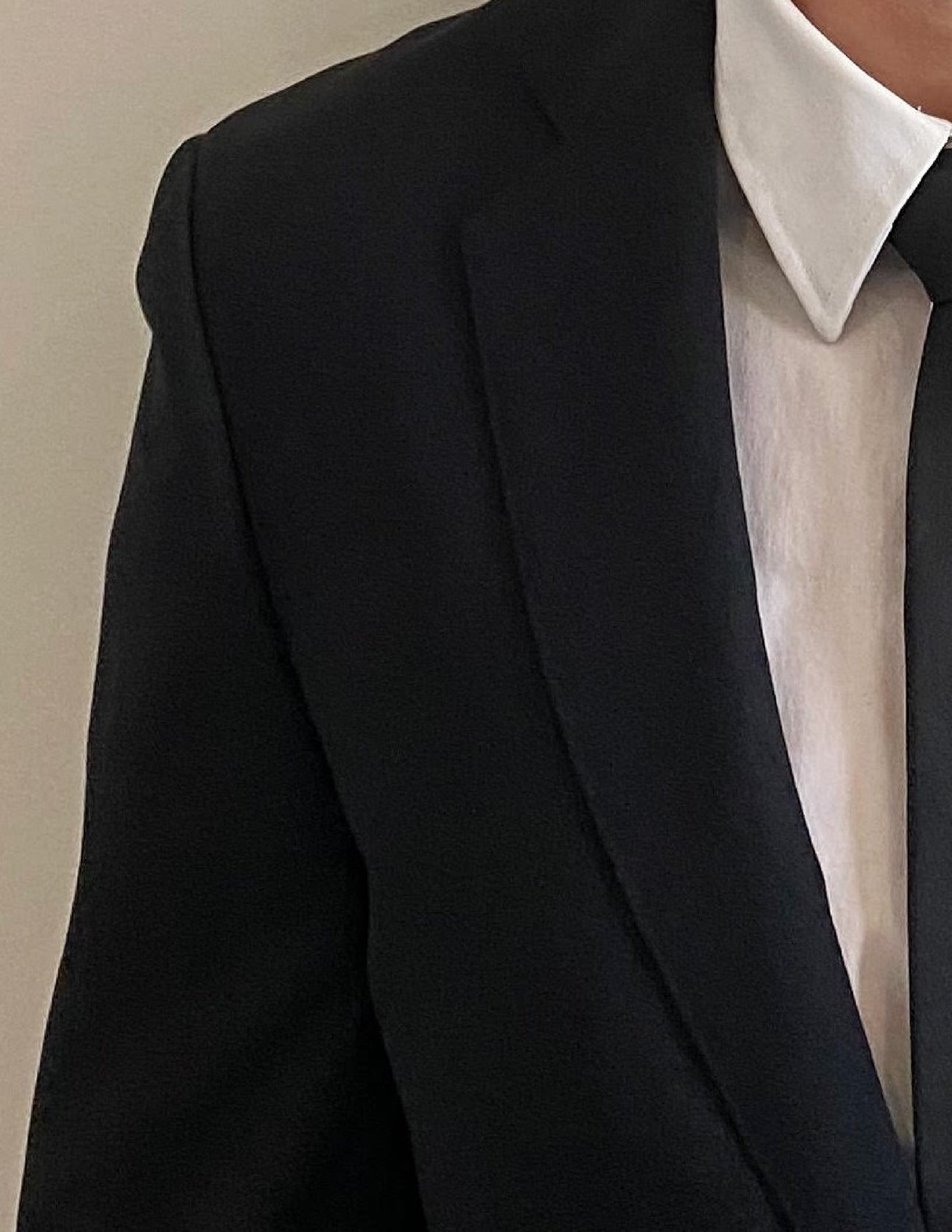}}]{Wen Jun Lim}
(Graduate Student Member, IEEE) received the Bachelor of Engineering (Telecommunication) and Master of Engineering Science from University of Malaya, Malaysia, in 2014 and 2017 respectively. He is currently pursuing the Ph.D. degree with the Centre for IoT and Telecommunications, The University of Sydney. His major research interests are channel coding and wireless communications.
\end{IEEEbiography}

\begin{IEEEbiography}[{\includegraphics[width=1in,height=1.25in,clip,keepaspectratio]{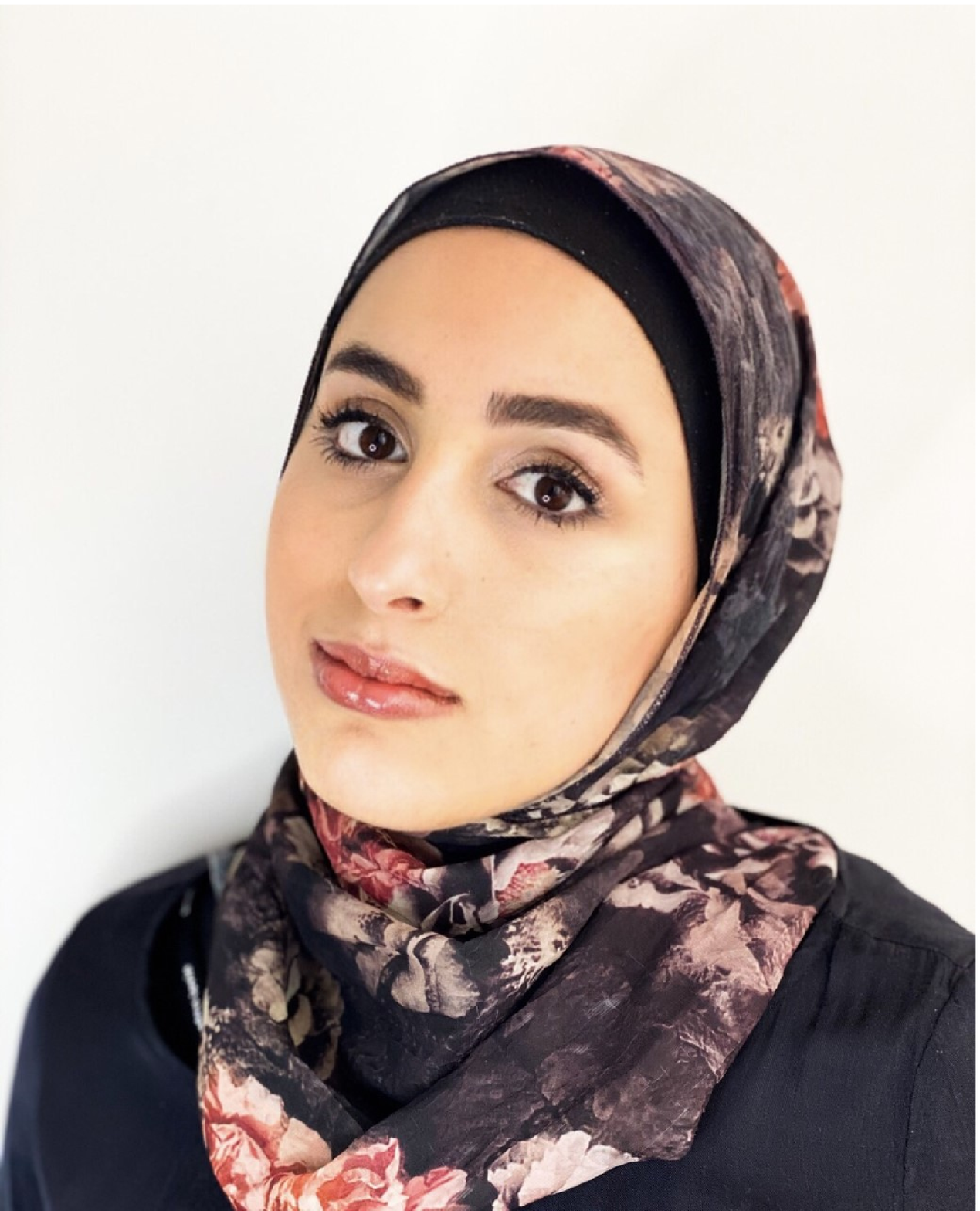}}]{Rana Abbas}
(Member, IEEE) received the B.Sc. degree from the University of Balamand, Lebanon, in 2012, and the M.E. and Ph.D. degrees in electrical engineering from The University of Sydney, in 2013 and 2018,  respectively. She is currently a Researcher at the Centre of IoT and Telecommunications, The University of Sydney, and an Innovation Engineer at the Engineering Excellence Group of Laing O’Rourke, Australia. Her research interests include channel coding, random access, information theory, machine-type communications, IIoT, and LiDAR. She was a recipient of the Australian Postgraduate Awards’ Scholarship and the Norman 1 Price Scholarship from the Centre of Excellence in Telecommunications, School of Electrical and Information Engineering, The University of Sydney. She was also a recipient of the Best Paper Award at the IEEE PIMRC, 2018.
\end{IEEEbiography}

\begin{IEEEbiography}[{\includegraphics[width=1in,height=1.25in,clip,keepaspectratio]{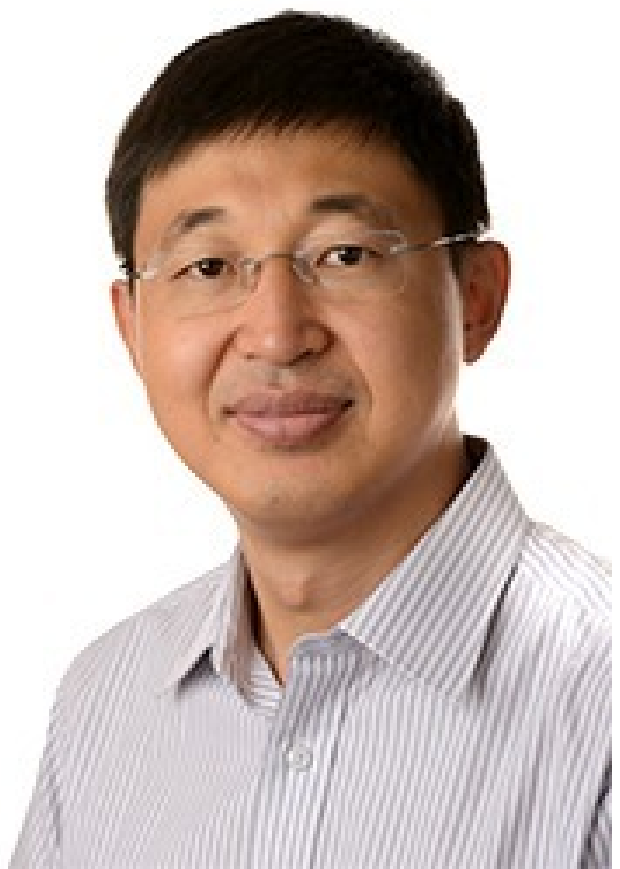}}]{Yonghui Li}
(Fellow, IEEE) received the Ph.D.degree from the Beijing University of Aeronautics and Astronautics, Beijing, China, in November 2002. Since 2003, he has been with the Centre of Excellence in Telecommunications, The University of Sydney, Sydney, NSW, Australia, where he is currently a Professor and the Director of the Wireless Engineering Laboratory, School of Electrical and Information Engineering. His current research interests are in the area of wireless communications, with a particular focus on MIMO, millimeter-wave communications, machine-to-machine communications, coding techniques, and cooperative communications. He holds a number of patents granted and pending in these fields. Dr. Li was a recipient of the Australian Queen Elizabeth II Fellowship in 2008 and the Australian Future Fellowship in 2012. He received the Best Paper Awards from the IEEE International Conference on Communications (ICC) 2014, the IEEE Personal, Indoor and Mobile Radio Communications (PIMRC) 2017, and the IEEE Wireless Days Conferences (WD) 2014. He has served as a Guest Editor for several IEEE journals, such as IEEE JSAC, IEEE Communications Magazine, IEEE IoT Journal, and IEEE Access. He is also an Editor of IEEE Transaction on Communications and IEEE Transaction on Vehicular Technology.
\end{IEEEbiography}

\begin{IEEEbiography}[{\includegraphics[width=1in,height=1.25in,clip,keepaspectratio]{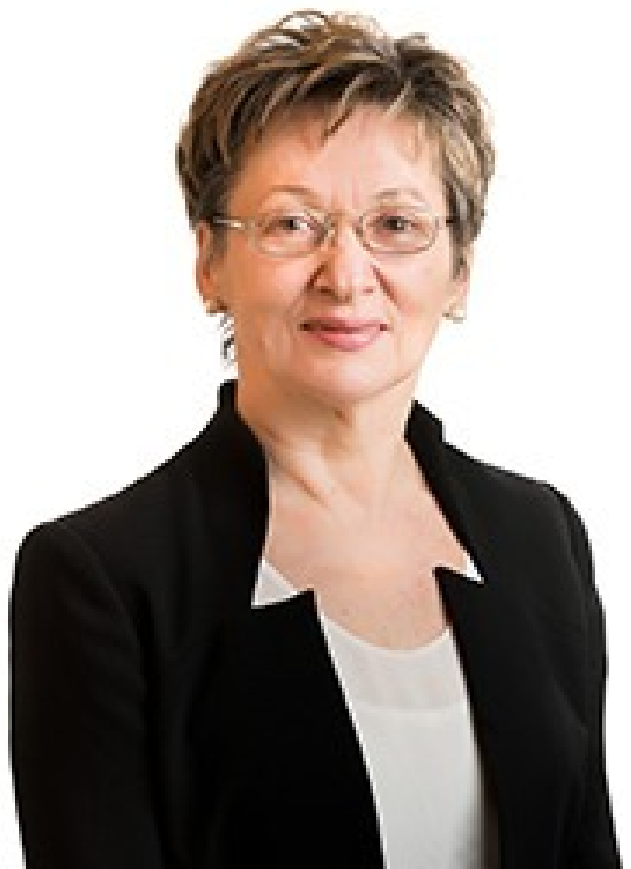}}]{Branka Vucetic}
(Life Fellow, IEEE)  is currently an ARC Laureate Fellow and the Director of the Centre of Excellence for IoT and Telecommunications, The University of Sydney, Sydney, NSW, USA. Her current research work is in wireless networks and the Internet of Things. In the area of wireless networks, she works on ultra-reliable low-latency communications (URLLC) and system design for millimeter-wave frequency bands. In the area of the Internet of Things, she works on providing wireless connectivity for mission-critical applications. Dr. Vucetic is a Fellow of the Australian Academy of Technological Sciences and Engineering and the Australian Academy of Science.
\end{IEEEbiography}

\begin{IEEEbiography}[{\includegraphics[width=1in,height=1.25in,clip,keepaspectratio]{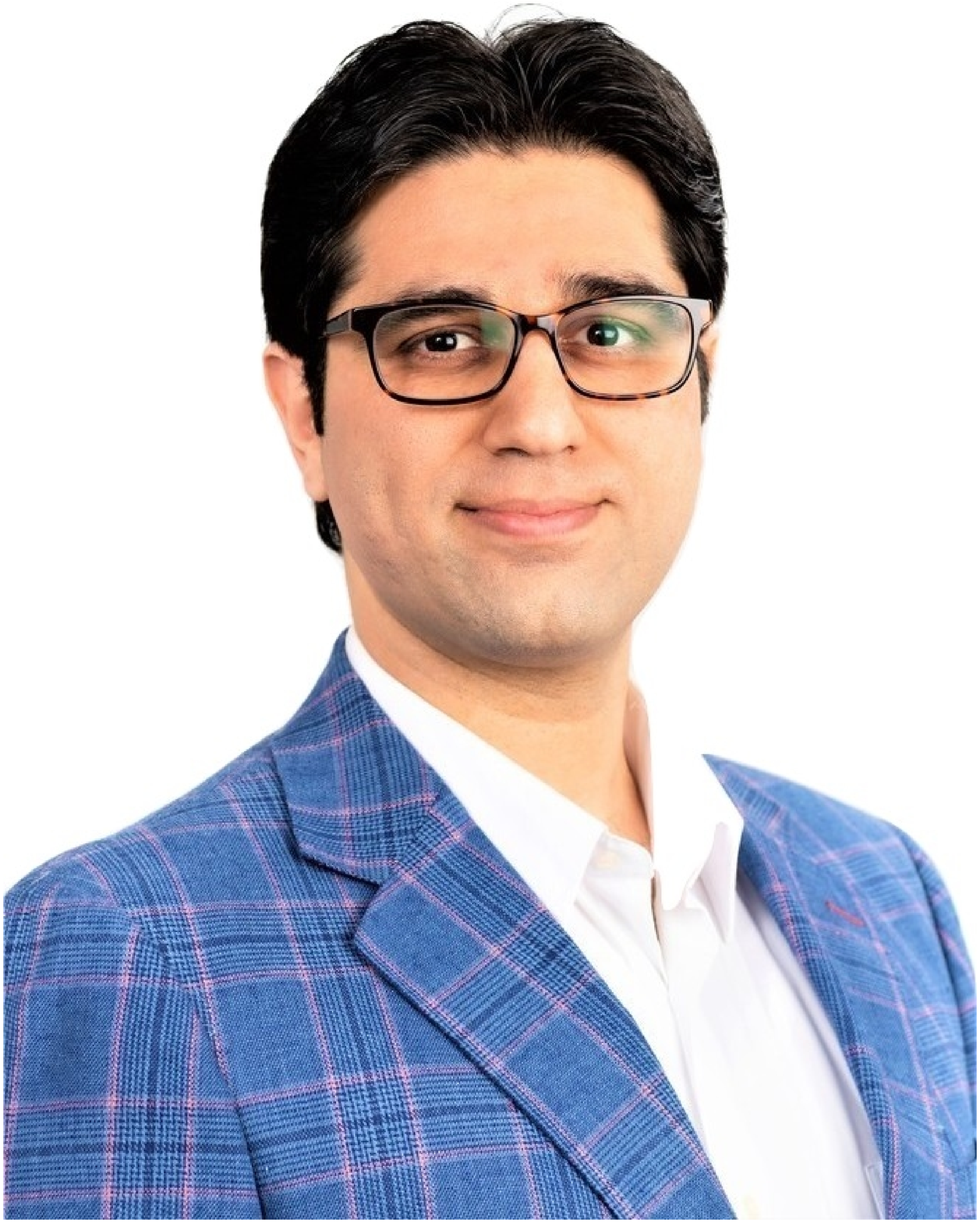}}]{Mahyar Shirvanimoghaddam}
(Senior Member, IEEE) received the B.Sc. degree (Hons.) from The University of Tehran, Iran, in 2008, the M.Sc. degree (Hons.) from Sharif University of Technology, Iran, in 2010, and the Ph.D. degree from The University of Sydney, Australia, in 2015, all in Electrical Engineering. He is currently a Lecturer with the Centre for IoT and Telecommunications, The University of Sydney. His research interests include coding and information theory, rateless coding, communication strategies for the Internet of Things, and information-theoretic approaches to machine learning. He is a fellow of the Higher Education Academy. He was selected as one of the Top 50 Young Scientists in the World by the World Economic Forum in 2018 for his contribution to the Fourth Industrial Revolution. He received the Best Paper Award for the 2017 IEEE PIMRC, The University of Sydney Postgraduate Award and the Norman I Prize, and The 2020 Australian Award for University Teaching.
\end{IEEEbiography}

\end{document}